\documentclass{appolb}
\usepackage{epsfig}
\usepackage{amsmath}
\def\lsim{\raise0.3ex\hbox{$<$\kern-0.75em\raise-1.1ex\hbox{$\sim$}}}
\def\gsim{\raise0.3ex\hbox{$>$\kern-0.75em\raise-1.1ex\hbox{$\sim$}}}

\begin{document}
\title{Thermodynamics of strong interaction matter from lattice QCD and 
the hadron resonance gas model
\thanks{Presented at XXXI Max Born Symposium: Three Days of critical 
behavior in hot and dense QCD,
Wroc\l aw, June 14-16, 2013}%
}
\author{Frithjof Karsch
\address{Physics Department, Brookhaven National Laboratory, Upton, NY 11973, USA\\
and\\
Fakult\"at f\"ur Physik, Universit\"at Bielefeld, D-33615 Bielefeld, Germany}
}
\maketitle
\begin{abstract}
We compare recent lattice QCD calculations of higher order
cumulants of net-strangeness fluctuations with hadron resonance
gas (HRG) model calculations. Up to the QCD transition temperature 
$T_c=( 154\pm 9)$~MeV we find good agreement
between QCD and HRG model calculations of second and fourth order 
cumulants, even when subtle aspects of net-baryon number, strangeness
and electric charge fluctuations are probed. In particular, the fourth 
order cumulants indicate that also in the strangeness sector of QCD the 
failure of HRG model calculations sets in quite abruptly in the vicinity 
of the QCD transition temperature and is apparent in most observables 
for $T\ \gsim\ 160$~MeV.
\end{abstract}
  
\section{Introduction}

A critical point, which is the endpoint of a line of first order
phase transitions, has been postulated to exist in the QCD phase diagram 
at non-zero values of the baryon chemical potential $\mu_B$ \cite{Stephanov}. 
Hints for the existence of such a critical point came from 
lattice QCD calculations that used a reweighting of Monte Carlo
data generated at vanishing chemical potential to non-zero chemical potential
\cite{Katz}. The validity
of these results, however, have been challenged \cite{Ejirireweighting}. 
The existence of a critical point for moderate values of 
$\mu_B/T \sim {\cal O}(1)$ has also been questioned on the basis of results
obtained in lattice QCD calculations with imaginary values of the chemical
potential \cite{deForcrand}.
Further information on the existence of a critical point and estimates for
its location can be obtained from higher order cumulants calculated at 
vanishing baryon chemical potential. Cumulants are the expansion parameters 
of Taylor series for
basic thermodynamic quantities, e.g. the pressure, baryon number density or
susceptibilities \cite{Allton,Gavai}. Estimates for the radius of convergence
of these series and the location of the critical endpoint at 
non-zero baryon number density can be obtained from ratios of higher order
cumulants of net-baryon number fluctuations, although it is not guaranteed
that these estimators converge rapidly \cite{Wagner}. Nonetheless, some 
estimates for the location of the critical point have been obtained in this 
way from calculations with unimproved staggered fermion actions on coarse 
lattices \cite{GG}.

Higher order cumulants of conserved charge fluctuations, i.e.
fluctuations of net-baryon number, electric charge and strangeness,
play a central role in the search for the critical endpoint and the 
exploration of the QCD phase diagram at vanishing and non-vanishing 
baryon chemical potential in general \cite{mborn2011}.
At non-vanishing baryon chemical potential quadratic fluctuations
of net-baryon number will diverge in the vicinity of a critical point
which makes them well suited also for the experimental search for the
existence of such a prominent landmark in the QCD phase diagram \cite{STAR}.

\begin{figure}[t]
\begin{center}
\includegraphics[width=8.5cm]{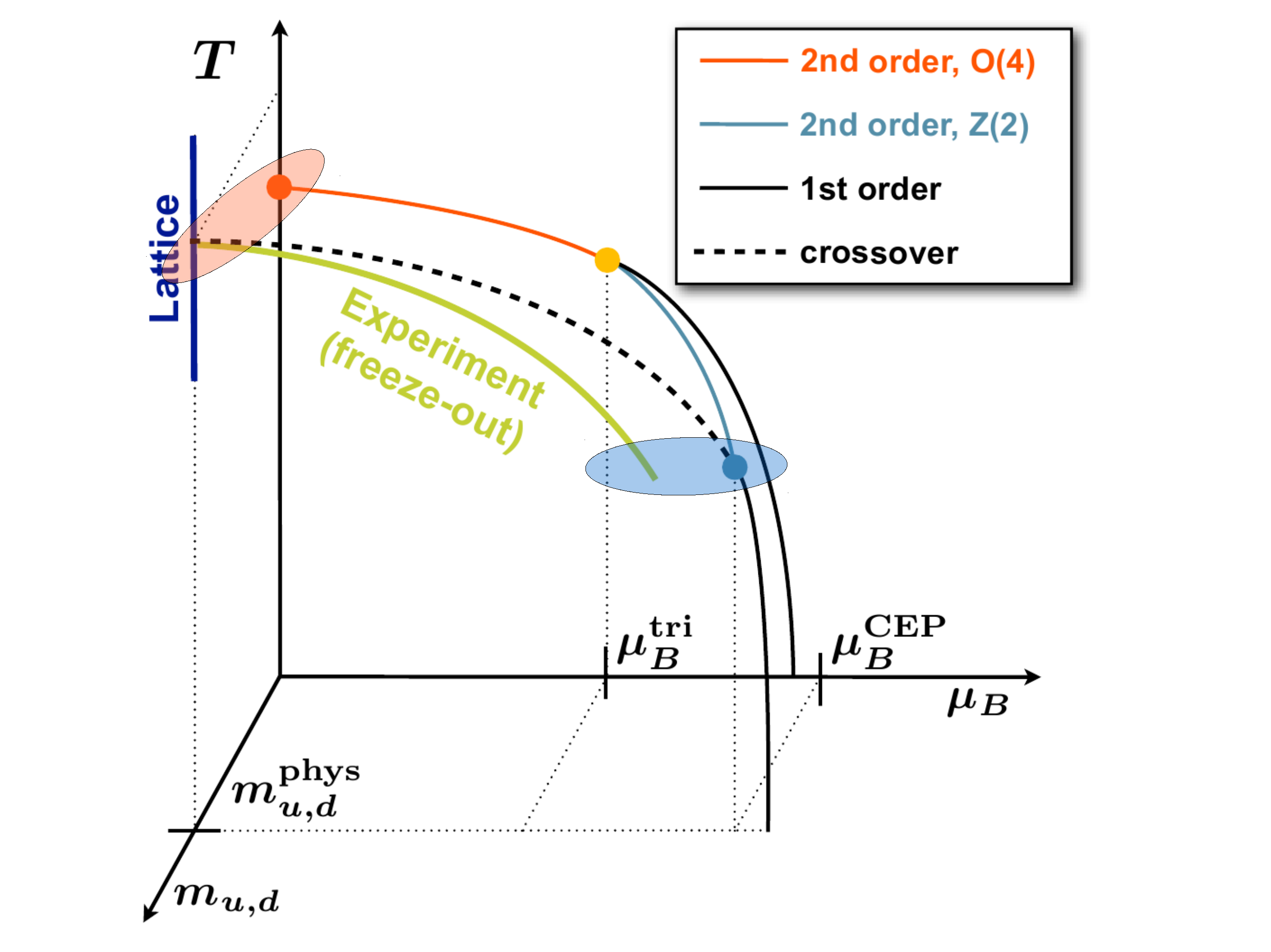}
\end{center}
\caption{Conjectured phase diagram of QCD in the space of temperature, 
baryon chemical potential and light quark mass.  
\label{fig:phase}}
\end{figure}

Irrespective of the existence of a critical endpoint, 
ratios of higher order cumulants of charge fluctuations provide detailed
information on properties of the different phases of strong interaction
matter. They signal the change in
relevant degrees of freedom in different phases 
\cite{Ejiri} and they also are sensitive probes for
the occurrence of a chiral phase transition at vanishing as well as 
non-vanishing baryon chemical potential. While in the latter case third 
order cumulants will diverge in the chiral limit, at vanishing chemical
potential only sixth order cumulants will for the first time show 
divergent behavior \cite{mborn2011}.
The critical regions, eventually probed through the calculation
of higher order cumulants, are illustrated in Fig.~\ref{fig:phase}.

Below the QCD transition temperature cumulants of net-charge fluctuations 
are known to agree quite well with hadron resonance gas (HRG) model 
calculations \cite{Ejiri,Tawfik}. This has been analyzed in 
detail for quadratic fluctuations of net-charges \cite{Wup,hotQCDHRG}.
If this gets confirmed also for higher order 
cumulants, 
it puts severe
constraints on estimates for the location of a critical point based on 
Taylor series expansions. In this conference contribution we 
will focus on cumulants involving
strangeness degrees of freedom \cite{strangeness}.
We want to discuss net-strangeness fluctuations and correlations with
net-baryon number fluctuations and will quantify to what extent
higher order cumulants, involving strangeness fluctuations, agree with
hadron resonance gas model calculations. We will show that strangeness 
fluctuations start deviating from HRG model calculations at or close to the 
QCD crossover temperature. Using also information on the modification of 
thermal strange meson correlation functions we argue that this suggests the
disappearance of strange hadronic bound states in the high temperature phase
of QCD.

\section{Confronting QCD results on higher order charge fluctuations with Hadron Resonance Gas model calculations}

Fluctuations of conserved charges and the correlation among moments
of net-charge fluctuations can be derived from 
the logarithm of the QCD partition function, which defines the pressure, $p$, 
\begin{equation}
\frac{p}{T^4} \equiv \frac{1}{VT^3}\ln Z(V,T,\mu_B,\mu_S,\mu_Q) \; .
\label{pressure}
\end{equation}
Taking derivatives with respect to chemical potentials for
baryon number ($\mu_B$), strangeness ($\mu_S$) and electric charge ($\mu_Q$)
evaluated at $\vec{\mu}=(\mu_{B},\mu_{Q},\mu_{S})=0$, we obtain higher order
cumulants of charge fluctuations $(\chi_n^X)$ and correlations 
$(\chi_{nm}^{XY})$ among moments of these charge fluctuations,
\begin{eqnarray}
\chi_n^X  = \left. \frac{\partial^n p/T^4}{\partial\hat{\mu}_X^n}  
\right|_{\vec\mu =0} & , &  
\chi_{nm}^{XY}  = \left. 
\frac{\partial^{n+m}p/T^4}{\partial \hat{\mu}_X^n \partial \hat{\mu}_Y^m}\right|_{\vec\mu =0} \; .
\label{obs}
\end{eqnarray}
Here we use the notation $\hat\mu_X \equiv \mu_X/T$ and $X,\ Y= B,\ Q,\ S$.

We will compare results for fluctuations and correlations defined by
Eq.~(\ref{obs}) with hadron resonance gas model  \cite{PBM} calculations.
The partition function of the HRG model can be split into mesonic and
baryonic contributions,
\begin{eqnarray}
\frac{p^{HRG}}{T^4} \hspace{-2mm}
&=&\frac{1}{VT^3}\sum_{i\in\;mesons}\hspace{-3mm} 
\ln{\cal Z}^{M}_{M_i}(T,V,\mu_Q,\mu_S)
\nonumber \\
&&\hspace{-3mm} +\frac{1}{VT^3}
\sum_{i\in\;baryons}\hspace{-3mm} \ln{\cal Z}^{B}_{M_i}(T,V,\mu_B,\mu_Q,\mu_S)\; ,
\label{ZHRG}
\end{eqnarray}
where the partition function for mesonic ($M$) or baryonic ($B$) particle
species $i$ with mass $M_i$ is given by
\begin{eqnarray}
\ln{\cal Z}^{M/B}_{M_i}
&=& 
{{VT^3}\over
{2\pi^2}}d_i \left(\frac{M_i}{T}\right)^2
\sum_{k=1}^\infty (\pm 1)^{k+1} {1\over {k^2}} K_2({{k M_i/T}})
\nonumber \\
&~&\hspace{2.5cm}\times 
\exp\left(k(B_i\mu_B+Q_i\mu_Q+S_i\mu_S)/T\right)
\ .
\label{ZMB}
\end{eqnarray}
Here upper signs correspond to mesons and lower signs to baryons.
In the temperature range of interest to us the Boltzmann approximation,
which amounts to restricting the sums in Eq.~\ref{ZMB} to the $k=1$ term 
only, is a good approximation for all particle species, except for 
pions. We will use this approximation in the following discussion.

\begin{figure}[t]
\begin{center}
\hspace{-0.6cm}\includegraphics[width=6.8cm]{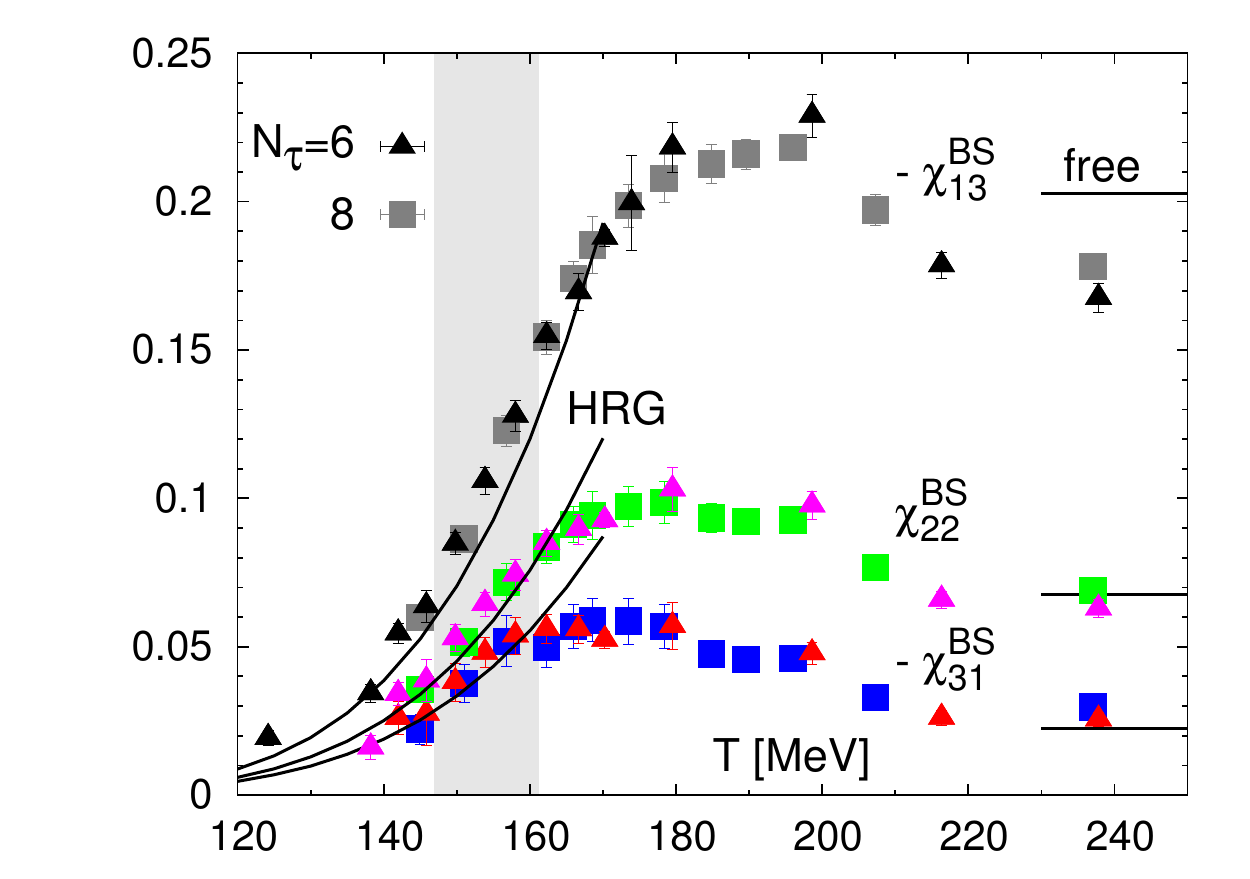}
\hspace{-0.6cm}\includegraphics[width=6.8cm]{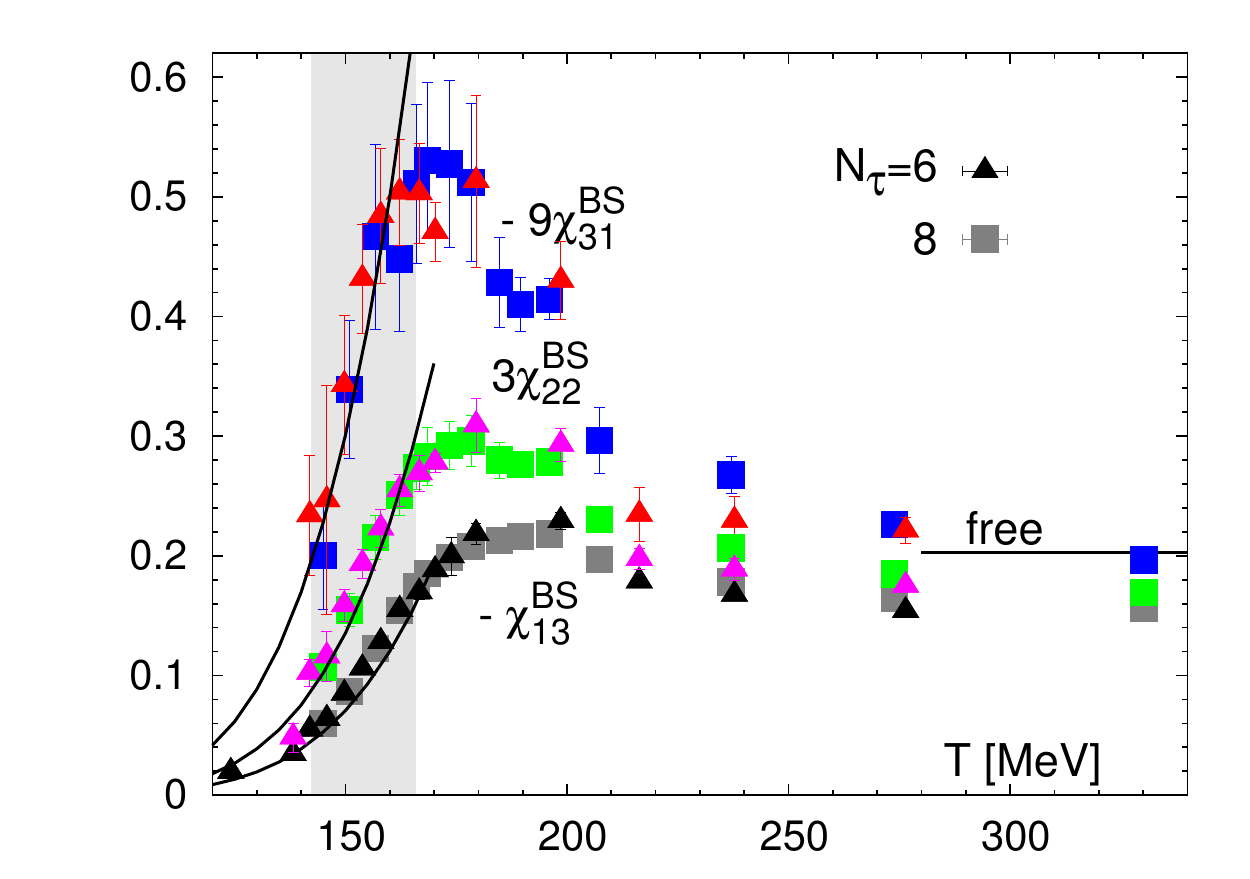}
\end{center}
\caption{Fourth order correlations of moments of net-baryon number
and net-strangeness fluctuations. Curves on the low temperature
side give HRG model results and the vertical lines on the high
temperature side give the ideal gas results. The grey band corresponds
to the crossover transition region $T_c=(154\pm 9)$~MeV \cite{hotQCDTc}
determined by the hotQCD Collaboration from the chiral susceptibility.
\label{fig:BS}}
\end{figure}

In Fig.~\ref{fig:BS} we show results for some $4^{th}$ order 
correlations of moments of net-baryon number and net-strangeness 
fluctuations, 
$\chi_{nm}^{BS}$, with $n+m=4,\ n>0$. 
In a gas of uncorrelated hadrons these correlations
receive contributions from baryons in the three different 
strangeness sectors, $|S|=1,\ 2$ and $3$. 
Within the HRG model approximation $\chi_{nm}^{BS}$
can be represented by a weighted sum of partial baryonic pressure contributions,
$P_{|S|=m,B}^{HRG}$, arising from the different strangeness sectors
\cite{strangeness},
\begin{equation}
\left( \chi_{nm}^{BS}\right)_{HRG} = 
P_{|S|=1,B}^{HRG} +2^m P_{|S|=2,B}^{HRG} + 3^m P_{|S|=3,B}^{HRG} \ , \ n>0\ .
\label{sectors}
\end{equation}
With increasing power $m$ of the strangeness moments the correlations 
$\chi_{nm}^{BS}$ thus give larger weight to multiple strange baryons,
i.e. an ordering $\chi_{31}^{BS}<\chi_{2}^{BS}<\chi_{13}^{BS}$ is naturally
expected.
As can be seen in the left hand part
of Fig.~\ref{fig:BS} this is indeed the case (even in the high temperature 
phase). Moreover, all three fourth order correlations $\chi_{nm}^{BS}$ 
agree quite well with HRG model calculations up to temperatures close to 
the QCD transition temperature. 

In the infinite temperature, ideal gas limit BS-correlations are 
expected to approach,
\begin{equation}
\left( \chi_{nm}^{BS}\right)_{free} 
= (-1)^m \frac{1}{3^n} 
\begin{cases}
1 & , \ n+m=2 \nonumber \\
\frac{6}{\pi^2} & , \ n+m=4 \\
\end{cases} \; .
\label{ideal}
\end{equation} 
This indeed seems to be a good approximation for temperatures 
$T\ \gsim\ 2T_c$ as can be seen from the right hand part
of Fig.~\ref{fig:BS} where we show
the fourth order BS-correlations rescaled such that their high temperature 
ideal gas limits coincide. 
At high temperature the baryon-strangeness
correlations thus suggest that the degrees of freedom carrying strangeness
are weakly interacting quasi-particles with quark quantum numbers, 
$B=1/3$, $S=-1$. 
However, in the temperature range $T\le T\ \lsim\ 2 T_c$ such a picture of 
weakly interacting quasi-particle clearly does not apply (see also
discussion in \cite{strangeness}).

\begin{figure}[t]
\begin{center}
\hspace{-0.6cm}\includegraphics[width=6.6cm]{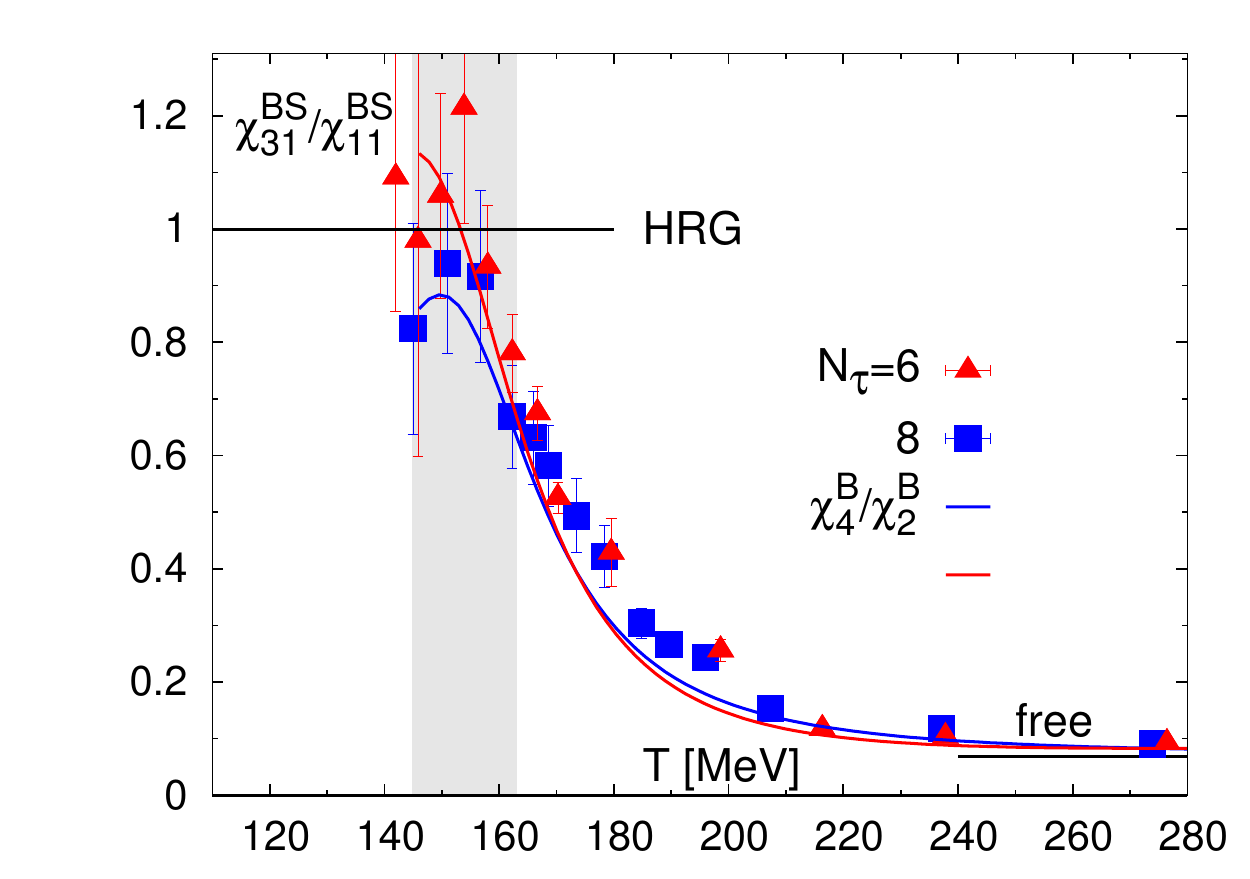}
\hspace{-0.6cm}\includegraphics[width=6.6cm]{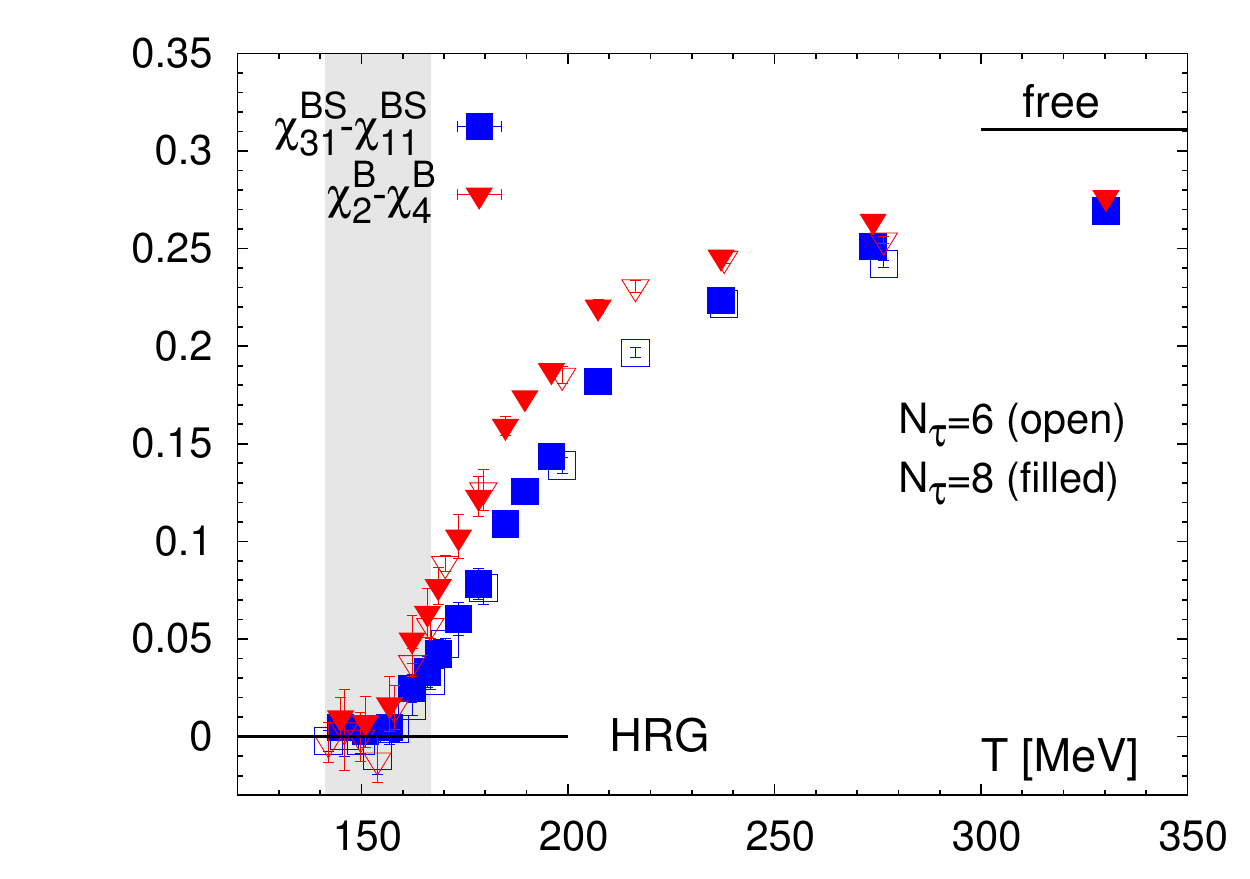}
\end{center}
\caption{Fourth order correlations of moments of net-baryon number
and net-strangeness fluctuations. Curves on the low temperature
side give HRG model results and the vertical lines on the high
temperature side give the ideal gas results. The grey band corresponds
to the crossover transition region $T_c=(154\pm 9)$~MeV \cite{hotQCDTc}
determined by the hotQCD Collaboration from the chiral susceptibility.
The curves in the left hand figure show fits to the ratio of 
net-baryon number cumulants, $\chi_4^B/\chi_2^B$, which receives i
contributions also from the non-strange baryon sector.
\label{fig:BSratio}}
\end{figure}

The agreement with HRG model calculations at low temperature can be 
probed in more detail by 
comparing second and fourth order BS-correlations that have identical
behavior at low temperature but widely different ideal gas limits.
Similar to the ratio of fourth and second order baryon number 
susceptibilities, $\chi_4^B/\chi_2^B$, which had been introduced
to probe the baryon number carrying degrees of freedom at low and 
high temperature \cite{Ejiri} one may consider 
$\chi_{31}^{BS}/\chi_{11}^{BS}$. This ratio is sensitive to the
strangeness carrying, baryonic degrees of freedom \cite{Koch}. Instead of 
this ratio the corresponding difference $\chi_{31}^{BS} -\chi_{11}^{BS}$
has been presented in Ref.~\cite{strangeness}. We show both variants
of this observable in Fig.~\ref{fig:BSratio}. We note that the differences
$\chi_2^B-\chi_4^B$ and $v_1 \equiv \chi_{31}^{BS} -\chi_{11}^{BS}$ show
similar behavior and drop rapidly at or close to the chiral transition
temperature. This has been confirmed in calculations using a different 
fermion discretization scheme \cite{Bellwied}. The temperature dependence of
these differences of second and fourth order cumulants is reminiscent of that 
of an order parameter. However, it should be clear that these observables 
are not 'order parameters' in the literal sense. Even in the chiral limit 
they would not vanish exactly below the QCD phase transition temperature, 
i.e. in the hadronic phase. In fact, in the vicinity of the chiral phase 
transition temperature the second and fourth order cumulants receive different
non-analytic contributions that are proportional to universal $O(4)$-scaling 
functions \cite{Engels}. This enforces deviations from simple HRG  
behavior in the vicinity of the QCD phase transition.

\begin{figure}[t]
\begin{center}
\includegraphics[width=6.4cm]{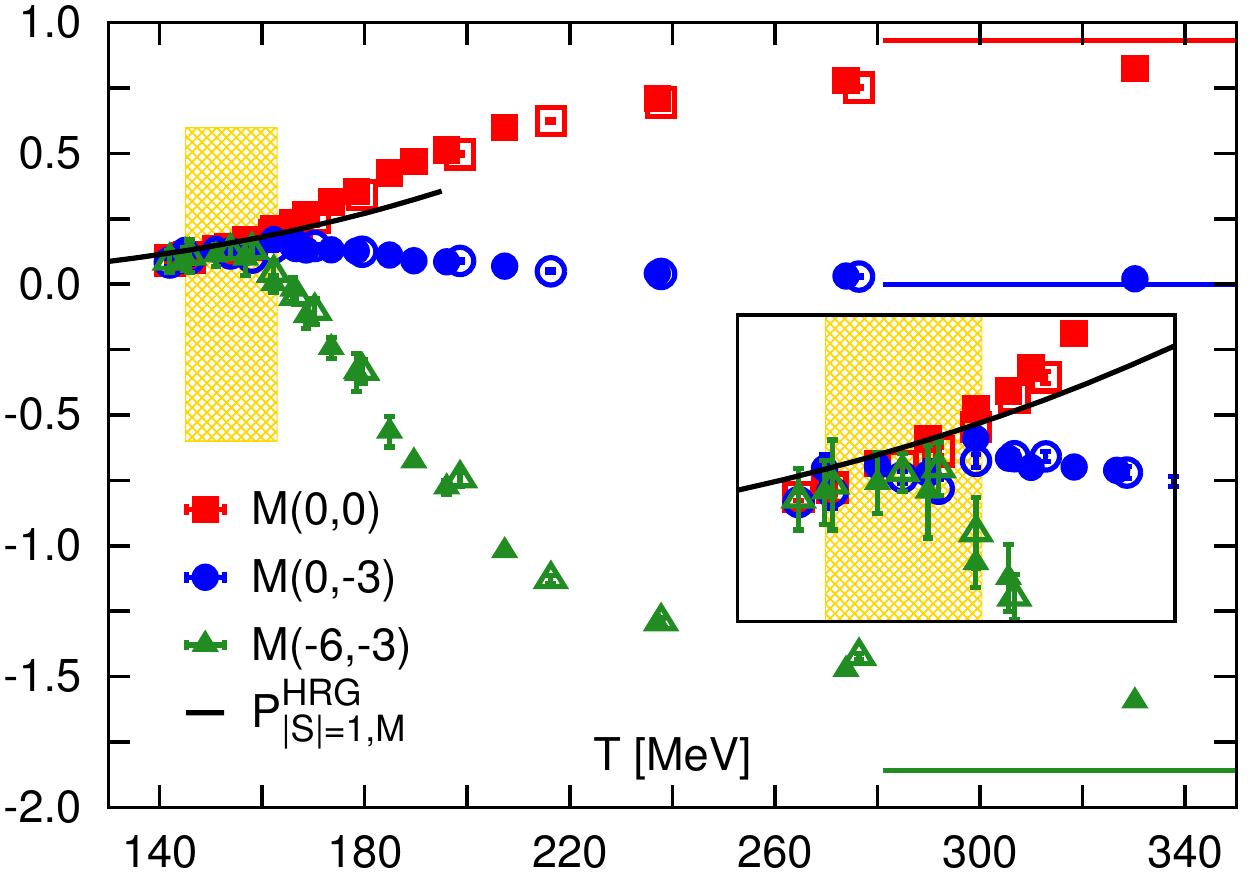}
\end{center}
\caption{Different combinations of net-strangeness fluctuations and
correlations with net-baryon number fluctuations that yield the
same strange meson contribution to the pressure in a
gas of uncorrelated hadrons. The yellow band indicates the temperature
range of the QCD crossover transition, $T_c=(154\pm 9)$~MeV. The black
line gives the pressure of strange mesons in a hadron resonance gas model.
The horizontal lines show the ideal gas well for the observables
$M(c_1,c_2)$ introduced in Eq.~\ref{meson}.
\label{fig:meson}}
\end{figure}

The BS-correlations discussed above are all sensitive to the strange 
baryon sector of QCD.
In order to become sensitive also to the strange meson sector
we construct observables that give the mesonic contribution to the
pressure in an uncorrelated hadron resonance gas. We can do so by
including also cumulants of strangeness fluctuation \cite{strangeness},
$\chi_n^S$, from which we subtract the baryon contribution by using 
suitable combinations of BS-correlations. This gives us quite some freedom.
We introduce 
\begin{equation}
M(c_1,c_2) = \chi_2^S -\chi_{22}^{BS} + c_1 v_1 + c_2 v_2\; ,
\label{meson}
\end{equation} 
where $c_1$ and $c_2$ are free parameters, $v_1$ has been introduced above 
and $v_2 = \frac{1}{3} (\chi_2^S - \chi_4^S ) - 2 \chi_{13}^{BS} -
4 \chi_{22}^{BS} - 2 \chi_{31}^{BS}$.  Similar to $v_1$ also $v_2$ is a 
combination of susceptibilities (independent from $v_1$) that vanishes in 
a gas of uncorrelated hadrons. 
We show in Fig.~\ref{fig:meson} this observable for three different choices
of $c_1$ and $c_2$. As can be seen at low temperatures, $T\lsim 160$~MeV,
and irrespective of the choice of $c_1,\ c_2$ the mesonic observables
$M(c_1,c_2)$ agree well with the strange meson contribution to 
the pressure of a HRG.

We thus conclude that at or close to the QCD transition temperature the HRG 
model breaks down as a description of mesonic and  baryonic degrees of freedom
in strong interaction matter.

\section{Strange Screening Masses}

\begin{figure}[t]
\begin{center}
\includegraphics[width=7.8cm]{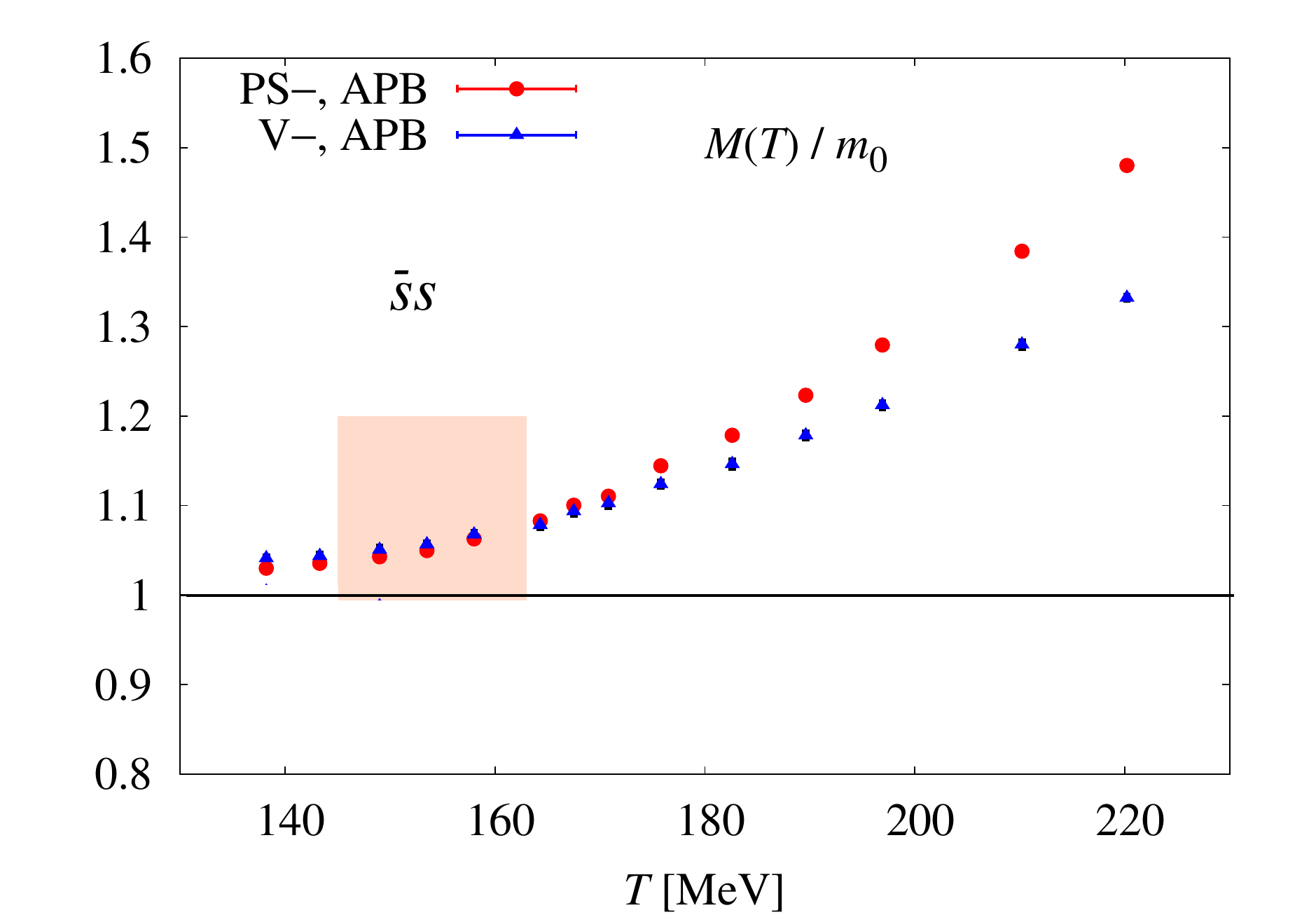}
\end{center}
\caption{Screening mass ($M(T)$) in units of the corresponding zero
temperature meson masses ($m_0$) extracted from spatial strange meson 
correlation functions in the pseudo-scalar (PS) and vector (V)
channels. Calculations are based on an analysis of gauge configuration
generated by the hotQCD Collaboration in (2+1)-flavor QCD using the HISQ 
action \cite{hotQCDTc}. The yellow band indicates the temperature
range of the QCD crossover transition, $T_c=(154\pm 9)$~MeV.
\label{fig:ss}}
\end{figure}

The analysis of various cumulants involving moments of net-strangeness
fluctuations presented in the previous section shows that the agreement 
with HRG model calculations breaks
down at temperatures close to the chiral transition temperature. 
Of course, deviations from the simple HRG results can also have
different origin and may, for instance, result from thermal modifications
of the hadron spectrum itself. First calculations of thermal meson spectral
functions performed in a quark mass regime corresponding to $\bar{s}s$-meson
states indeed suggest that such states may still exist in the QGP 
\cite{Hatsuda}. However, so far these calculations have only been performed 
in quenched QCD; the influence of screening due to dynamical quark degrees
of freedom may well lead to an earlier melting of strange meson states. 
In order to gain further insight into this question
we have analyzed spatial correlation functions of strange mesons \cite{Maezawa}.
These correlation functions have a representation in terms of finite 
temperature spectral functions, $\rho(\omega,p_z,T)$, although the 
contribution of resonance peaks is not directly evident in these correlators,
\begin{equation}
C(z,T) = \int_0^\infty \frac{2d\omega}{\omega} \int_{-\infty}^\infty dp_z e^{i p_z z}
\rho(\omega,p_z,T)
\;.
\end{equation}
The large distance behavior of these correlation functions yield 
temperature dependent screening masses, $M(T)$. In the infinite 
temperature limit $M(T)/T$ approaches twice the lowest Matsubara 
frequency reflecting the propagation of two uncorrelated quarks. 
In Fig.~\ref{fig:ss} we show the screening mass in pseudo-scalar and 
vector channels of strange quark-antiquark states. It is obvious that 
the screening masses show a strong temperature dependence already at
temperatures close to the QCD transition temperature. In fact, already
in the crossover region to the high temperature phase deviations from
the zero temperature $\bar{s}s$-meson mass, $m_0$, are about 5\%. In the
case of charmonium states, which are known to melt at about $1.5\ T_c$
\cite{Ding} (or earlier) the finite temperature screening masses deviate
from the zero temperature $J/\psi$ or $\eta_c$ masses only by about 2\%
\cite{Maezawa}.
This suggests that all strange meson and baryon states dissolve already 
at the QCD transition. 

\section{Conclusions}
The hadron resonance gas model provides a remarkably good description
of the thermodynamics of strong interaction matter in the low temperature
hadronic phase. Even subtle aspects of fourth order charge fluctuations
like the relative contributions of different strangeness sectors to bulk
thermodynamics
are well described by the HRG model. Its rapid breakdown in the crossover 
region is apparent also in the strange hadron sector and suggests that
strange mesons and baryons disappear at or close to the QCD transition.
 
\vspace*{-0.2cm}
\section*{Acknowledgments}
I gratefully acknowledge the many fruitful discussions and 
the collaboration I had with Krzysztof Redlich over a good fraction 
of the last 60 years that he is around. This would not have been
possible without the continuous support of both of our families.
This work was also supported in part by contract DE-AC02-98CH10886
with the U.S. Department of Energy.

\end{document}